\RequirePackage{fix-cm}
\documentclass[twocolumn]{svjour3}          
\smartqed  
\usepackage{graphicx}
\usepackage{amsmath}
\usepackage{subcaption}
\usepackage{color}

\newcommand{\red}[1]{\textcolor{red}{#1}}

\begin{document}

\title{Soft particles reinforce robotic grippers
}
\subtitle{robotic grippers based on granular jamming of soft particles}


\author{Holger G\"otz \and
        Angel Santarossa \and
        Achim Sack \and
        Thorsten P\"oschel \and
        Patric M\"uller
}


\institute{Holger G\"otz \and Angel Santarossa \and  Achim Sack \and Thorsten P\"oschel \and Patric M\"uller\at
Institute for Multiscale Simulations\\
Friedrich-Alexander-Universit\"at Erlangen-N\"urnberg\\
Cauerstra\ss{}e 3, 91058 Erlangen\\
Germany\\
\email{patric.mueller@fau.de}           
}

\date{Received: date / Accepted: date}

\maketitle

\begin{abstract}
Granular jamming has been identified as a fundamental mechanism for the operation of robotic grippers. In this work we show, that soft particles like expanded polystyrene beads lead to significantly larger gripping forces in comparison to rigid particles. In contradiction to naive expectation, the combination of jamming and elasticity gives rise to very different properties of the jammed phase, compared to hard-particle systems. This may be of interest also beyond the application in robotic grippers.

\end{abstract}

\section{Idea and state of the art of granular grippers}
\label{sec:intro}
An important and challenging problem in the context of robotics is to grip objects of various and sometimes unknown shape and surface properties with a single manipulator. Frequently, this problem is addressed by grippers that mimic hands with multiple fingers (e.g. \cite{monkman2006}). However, complex mechanics also requires complicated software that manages the interplay of sensory and actuating elements.
In 2010, revisiting some earlier ideas \cite{schmidt1978,Perovskii1980,burckhardt188}, Brown et al. \cite{Brown2010} suggested a different approach based on the jamming transition of granular materials. Here, granulate is loosely enwrapped into an elastic membrane. The filled membrane is then pressed onto an object such that the granulate flows around the object in a fluid-like manner. To grip the object, the air within the membrane is evacuated and eventually the granulate gets compressed and enters a solid-like state due to jamming. This solidification causes forces that can effectively grip and hold the object. More information on the underlying jamming transition in granular materials can be found, e.g., in \cite{jamming2001,nagel1998,biroli2007}. As described by Brown, this type of gripper mimics the limit of a robotic hand with infinitely many degrees of freedom \cite{Brown2010}. In contrast to ordinary robotic grippers, these degrees of freedom need not to be controlled explicitly. Instead, they automatically adapt to the object and they are collectively activated by evacuating the membrane. It has been shown in many subsequent publications \cite{Brown2010,amendPositivePressureUniversal2012,kapadiaDesignPerformanceNubbed2012,lichtPartiallyFilledJamming2018,jiangLearningHardwareAgnostic2012} that these grippers can operate for a wide range of shapes, fragile objects and even multiple objects at a time, without any reconfiguration or learning phases.

Brown et al. \cite{Brown2010} have shown that, depending on the properties of the object to be gripped, three different mechanisms are essential:
\begin{enumerate}
    \item geometric interlocking between the object and the solidified granulate,
    \item static friction between the object and the membrane, and
    \item suction effects, where the membrane seals the surface of the object air-tight.
\end{enumerate}
It was further shown, that the holding forces due to geometric interlocking and suction are significantly larger than those resulting from static friction \cite{Brown2010,lichtUniversalJammingGrippers2016}. Therefore most attempts to optimize the gripping performance were focused on interlocking and suction \cite{kapadiaDesignPerformanceNubbed2012,lichtPartiallyFilledJamming2018}. For those two mechanisms the strength of the particle material in its jammed state is crucial \cite{Brown2010}. Consequently, almost all works use particles from rather rigid materials like steel beads, glass beads, rice, salt or sugar.
In the current paper, we consider particles from soft materials instead of rigid ones and show that this leads to
a squeezing effect between the gripper and the object, i.e., the object is firmly pressed by the gripper due to its volume reduction (particles and membrane) when vacuum is applied. This effect significantly increases the static friction between the object and the membrane and turns gripping by static friction competitive with the other mechanisms.
We study this gripping force enhancement by means of experiments, particle simulations, and continuum mechanical theory. By that, we add one further mode of operation for granular gripping that makes it possible to tightly hold objects whose shape and material properties neither allow for suction effects nor geometric interlocking.


\section{Holding force enhancement due to soft particles\label{sec:enhancement}}
\begin{figure}
\centering
    \includegraphics[width=\columnwidth]{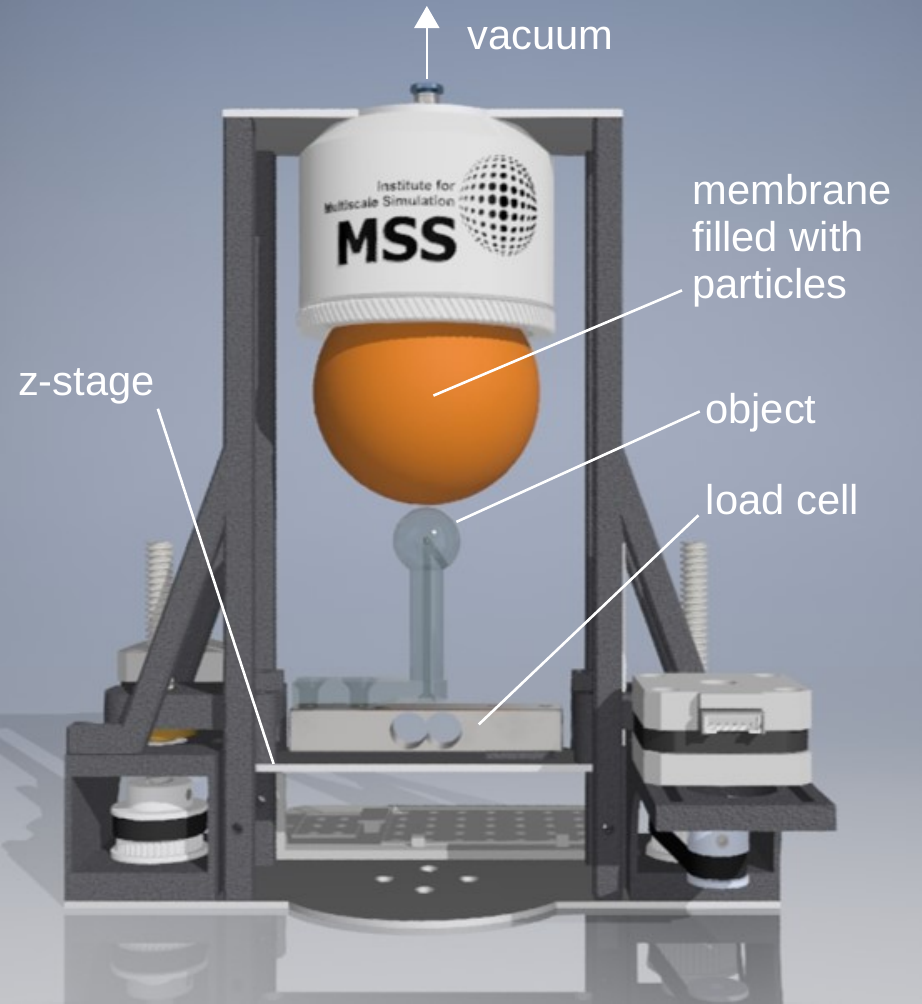}
    \caption{Experimental setup for automatic holding force measurements.\label{fig:expSetup}}
\end{figure}
In this section, we experimentally compared the performance of soft and rigid particles in a granular gripper by measuring the maximum holding force achieved in a gripping process. To that we reconsider the simple form of a granular gripper which has already been discussed by Brown et al. \cite{Brown2010}, where a single elastic and air-tight membrane of spherical shape (diameter $(73.0 \pm 0.5) \text{mm}$) is filled with granulate. The granulate we consider are glass beads of diameter $(4.0 \pm 0.3) \text{mm}$ and expanded polystyrene (EPS) beads of diameter $(4.2 \pm 0.5) \text{mm}$. The object to be gripped is a steel sphere of diameter $(19.99 \pm 0.01) \text{mm}$ whose surface was roughened to avoid suction effects between the membrane and the object. To measure the holding force, we use the apparatus shown in Fig.~\ref{fig:expSetup}. One measurement cycle consists of the following steps: first the membrane is pressurized to fluidize the enwrapped granulate, then, the z-stage is used to push the object into the gripper. Once the desired indentation depth is reached, the z-stage is stopped and after a relaxation phase the membrane is evacuated such that the pressure difference between outside and inside is $p_\text{vac}\approx 90 \text{kPa}$, which is the maximum pressure difference, that our vacuum-pump is able to achieve. We chose this value, as it has been experimentally shown that a higher vacuum pressure leads to higher holding forces \cite{lichtUniversalJammingGrippers2016}. After evacuating the membrane, the z-stage is moved downwards until the contact between the object and the gripper ceases. During the interval where the gripper is in touch with the object, the force acting between the object and the z-stage is measured.
\begin{figure}
\centering
    \includegraphics[width=\columnwidth]{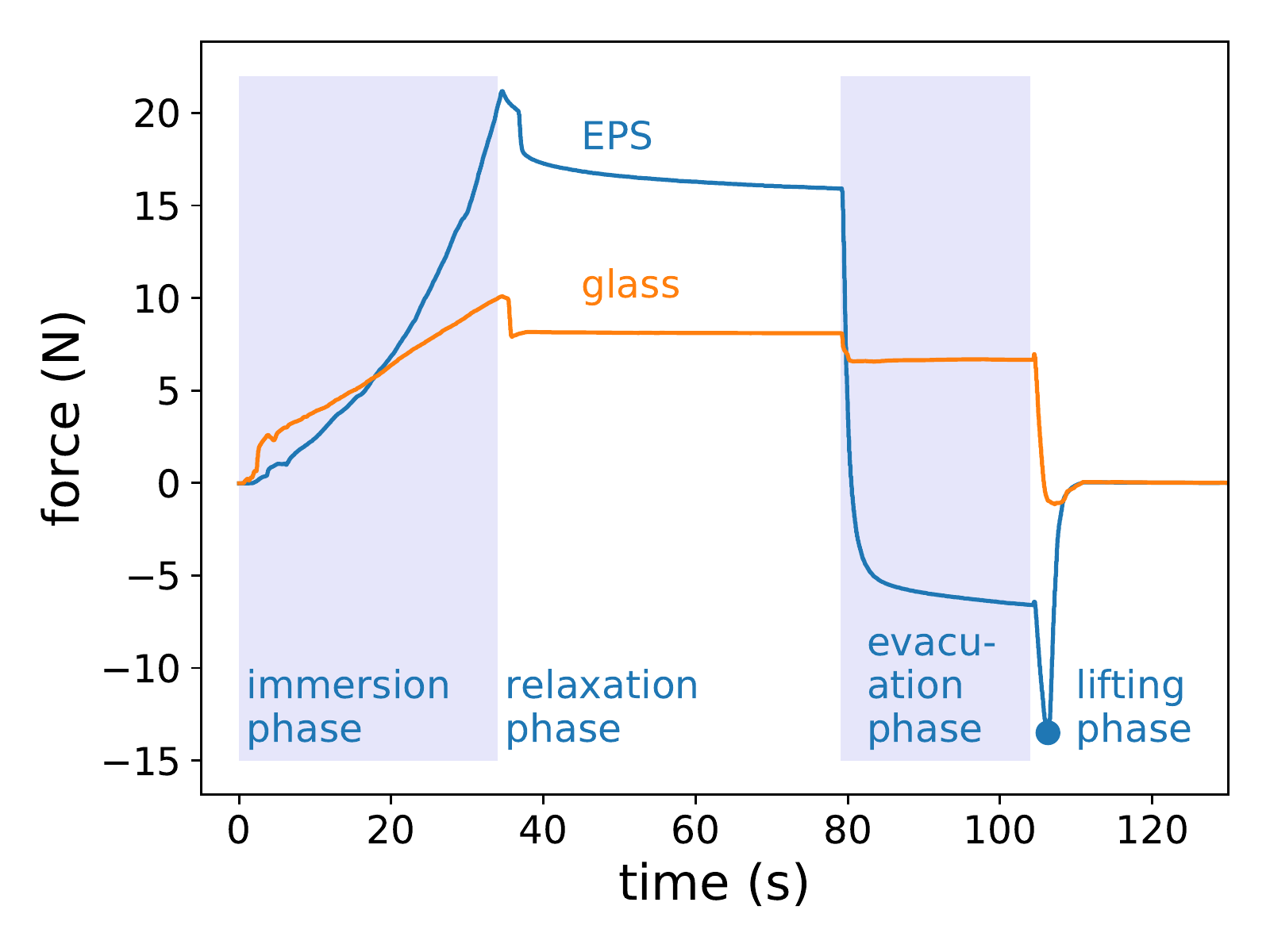}
    \caption{Force in $z$-direction as measured by the load cell shown in Fig.~\ref{fig:expSetup}. The blue curve shows the force signal for EPS beads, the orange curve the force signal for glass beads. The blue rectangles in the background and the blue text at the bottom indicate different phases of the gripping process (see text). The blue dot marks the maximum holding force ($\approx14\,\text{N}$) for the EPS beads.
\label{fig:forceSignals}}
\end{figure}
Fig. \ref{fig:forceSignals} shows the result for both, EPS beads and glass beads. Additionally, the characteristic phases of the gripping process are indicated for the measurement with both type of beads. First, the object is pushed into the granulate (immersion phase) which corresponds to an increasing force in $z$-direction. At the beginning of the relaxation phase, the force decreases suddenly due to the abrupt stopping of the z-stage. In the relaxation phase, the force in z-direction continuously saturates to a constant value due to the reorganization and deformation of the particles. For the EPS beads this process takes much longer as compared to the glass beads. The same holds true for the subsequent evacuation phase: in both cases, the force in z-direction decreases smoothly until a constant value is reached, as before, this process is much quicker for the glass particles due to their higher stiffness. Note that for the EPS beads the sign of the z-force attains negative values in the evacuation phase, which corresponds to a lifting force onto the object even during the evacuation phase where the z-stage is not moved at all. This is due to the softness of the EPS, which causes the whole gripper to shrink significantly while the evacuation is performed.
After the membrane is evacuated to the desired pressure, the z-stage is lowered and the force, as measured by the load cell (see Fig.~\ref{fig:expSetup}), decreases continuously in both cases, glass and EPS. Only now, the force for the glass beads decreases to negative values corresponding to a lifting force onto the object. If the $z$-stage is lowered down further, the gripper is no longer able to hold the object, the contact between the object and the gripper suddenly ceases and the measured force quickly vanishes. The minimum of the force in z-direction corresponds to the the maximum possible holding force of the system. Note that for the glass beads we observe a significantly lower maximum holding force compared to the EPS beads.


\begin{figure}
\centering
    \includegraphics[width=\columnwidth]{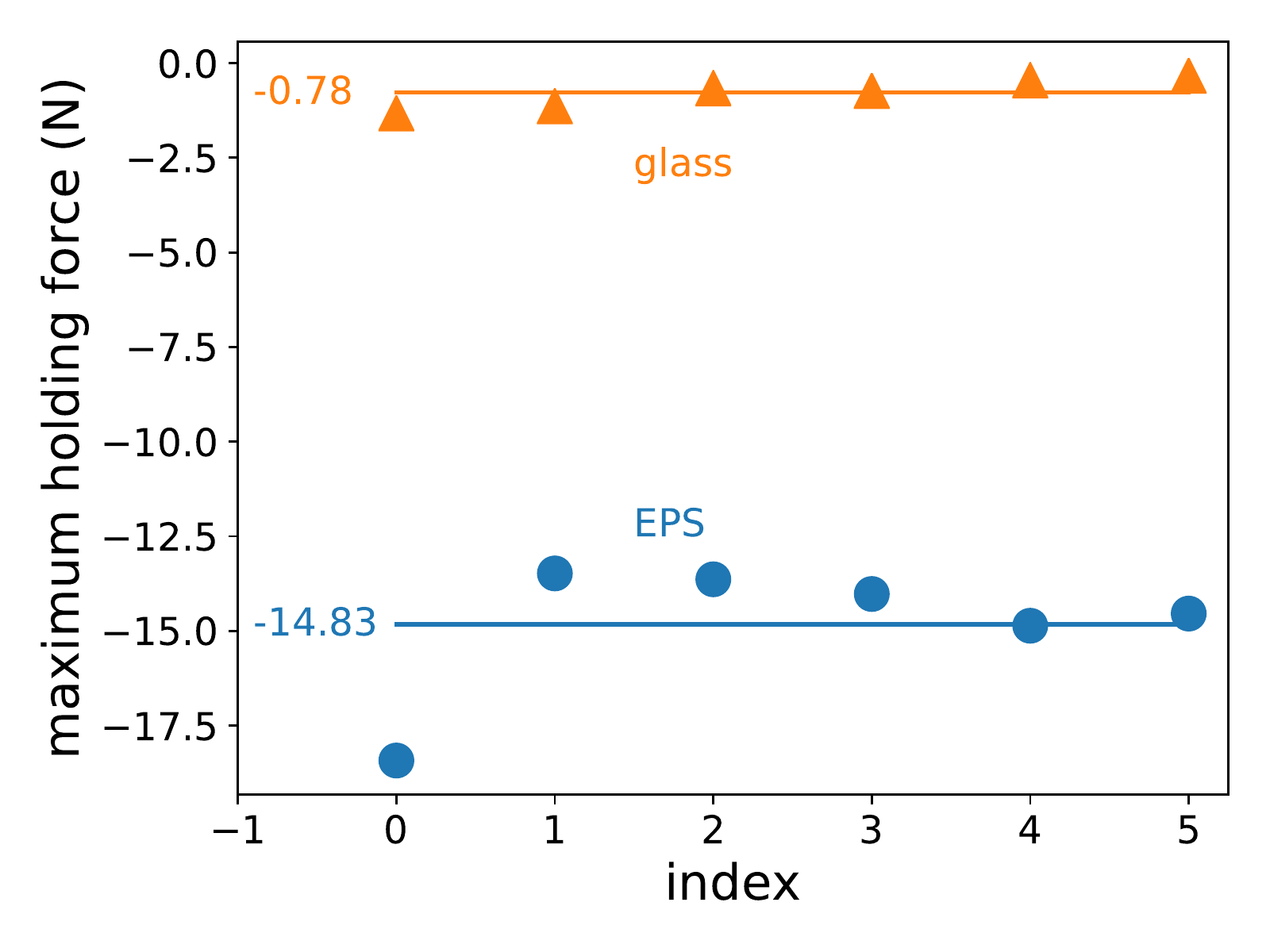}
    \caption{Maximal holding forces for glass beads (orange triangles) and EPS beads (blue circles) for six independent measurements. The solid lines indicate the mean value of the holding force which is noted left to the line.\label{fig:holdingForces}}
\end{figure}
Fig.~\ref{fig:holdingForces} shows the maximum holding forces for six independent experiments for both, glass beads and EPS beads. The data indicate, that on average, the holding force is increased by a factor of approximately 19 for the EPS beads.


\section{Operation principle for soft particle granular grippers}
\begin{figure}
\centering
    \includegraphics[width=0.5\columnwidth]{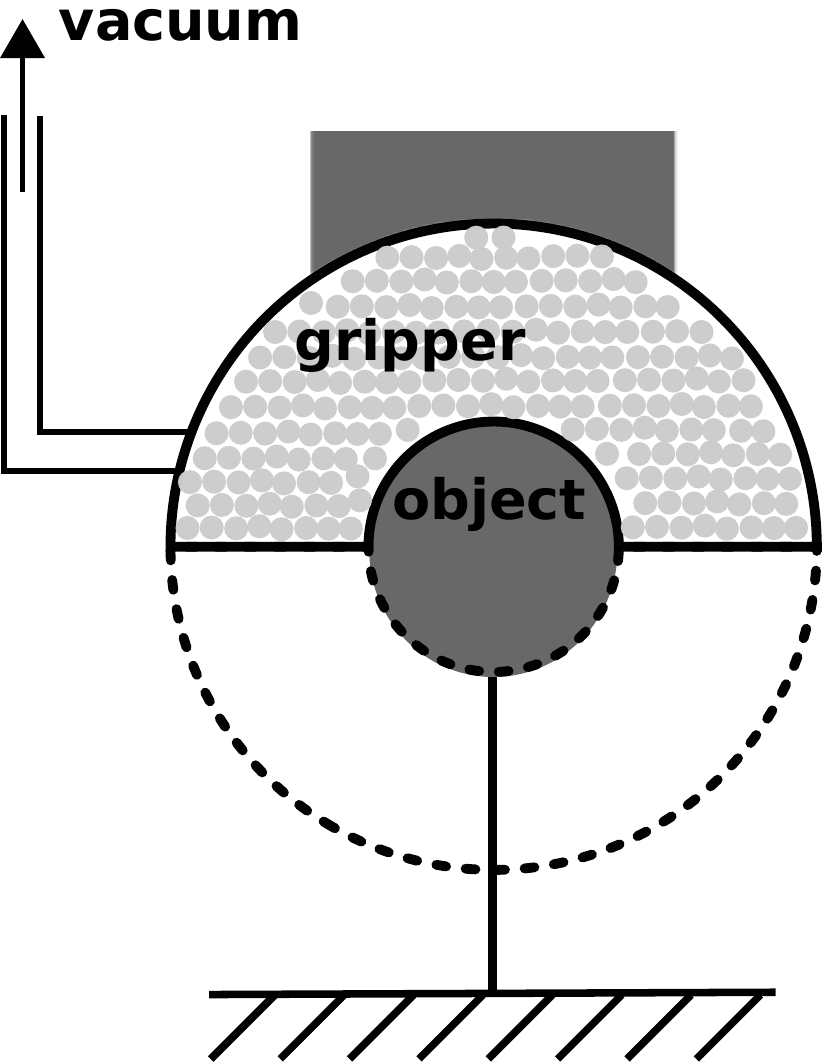}
    \caption{Sketch of the simplified model for the granular gripper.\label{fig:model}}
\end{figure}
In this section, we first explain the holding force enhancement by means of a simplified continuum mechanical model. In the second part, we use particle simulations to show that the results are valid also for the granular system.
\subsection{Continuum mechanical description}
\begin{figure}
\centering
    \includegraphics[width=0.95\columnwidth]{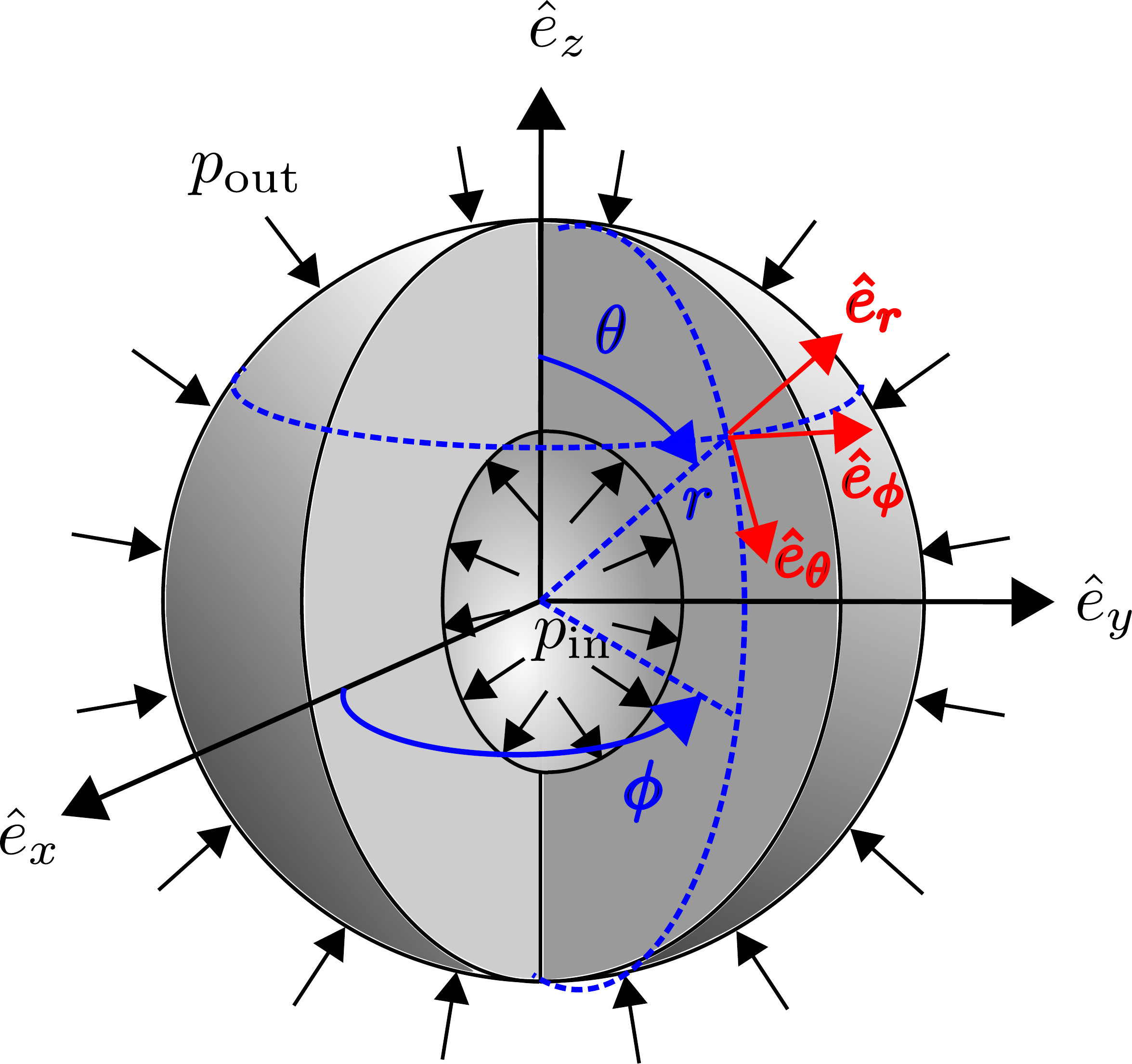}
    \caption{Sketch of a pressurized hollow sphere including mathematical notation.\label{fig:hollowSphere}}
\end{figure}

We consider the simplified model of the granular gripping system illustrated in Fig.~\ref{fig:model}. The membrane of the gripper encloses the volume of a hemispherical shell, i.e. the volume between two concentric hemispheres of radius $r_\text{in}$ and $r_\text{out}$. The object to be gripped is a sphere, such that only its upper hemisphere is in contact with the gripper and interlocking between the object and the gripper is not possible. We further assume, that there are no suction effects between the object and the membrane. In the experiment, this can be achieved, e.g., by using a porous object or an object with a rough surface. With these preconditions, the remaining gripping mechanism is due to friction between the membrane and the object. According to Coulomb's law, the friction is limited by the normal force $F_\text{n}$. In the following, we therefore derive the normal force the gripper exerts onto the object. This force results from the evacuation of the membrane. Due to the applied vacuum and the atmospheric pressure, there is a pressure difference $p_\text{vac}$ acting onto the surface of the membrane. To compute the resulting deformation of the gripper, we assume that the bulk of the granular material inside the membrane behaves like a linear elastic material with Young's modulus $E$ and Poisson's ratio $\nu$. Further, we now consider the full spherical shell (gripper part plus dashed line in Fig.~\ref{fig:model}) to simplify the boundary conditions. The problem we have to solve now corresponds to a pressurized hollow sphere where the pressure $p_\text{in}$ is acting onto the inner surface at $r = r_\text{in}$ and $p_\text{out}$ is acting onto the outer surface at $r = r_\text{out}$ (see Fig.~\ref{fig:hollowSphere}). For this spherically symmetric linear elastic problem, the displacement field $\vec{u}$ in the sphere reads (e.g. \cite{bower2009})
\begin{align}
\label{eq:dissplacement}
    \vec{u}(r)=&\frac{1}{2E(r_\text{out}^3-r_\text{in}^3)r^2}\cdot\\\nonumber
    &\left[\right.2\left(p_\text{in}r_\text{in}^3-p_\text{out}r_\text{out}^3\right)(1-2\nu)r^3+\\\nonumber
    &(p_\text{in}-p_\text{out})(1+\nu)r_\text{out}^3r_\text{in}^3\left.\right]\hat{e}_r.
\end{align}
Here we use spherical coordinates where $r$ is the radial coordinate and $\hat{e}_r$ is the unit vector in radial direction (see Fig.~\ref{fig:hollowSphere}).

We now assume, that the sphere we grip (see Fig.~\ref{fig:model}) is ideally rigid, i.e., it does not deform under load. In our simplified, spherically symmetric geometry, Fig.~\ref{fig:hollowSphere}, this corresponds to the case that the hollow sphere is filled by an undeformable smaller sphere. In this situation, we have to fulfill the constraint
\begin{equation}
\vec{u}(r=r_\text{in})=0.
\label{eq:constraint}
\end{equation}

For finite Young's modulus, the hollow sphere will be deformed according to Eq.~\eqref{eq:dissplacement} due to the external pressure. The evacuation of the membrane corresponds to $p_\text{vac} \equiv p_\text{out} = p_\text{in}$. To comply with the criterion Eq.~\ref{eq:constraint}, the object inside the hollow sphere has to react with a counterpressure $p_\text{c}$ such that it does not get compressed and, therefore, $p_\text{in}= p_\text{vac} + p_\text{c}$. If we solve Eq.~\ref{eq:constraint} for the counterpressure $p_\text{c}$ we find
\begin{equation}
p_\text{c} = p_\text{vac}\frac{\left(2-\left(\frac{r_\text{out}}{r_\text{in}}\right)^3\right)(2\nu-1)}{2(1-2\nu)+\left(\frac{r_\text{out}}{r_\text{in}}\right)^3(1+\nu)}.
\label{eq:pc}
\end{equation}
Naively one would expect the counterpressure to depend on the the Young's modulus of the material of the hollow sphere, however, this is not the case and $p_\text{c}$ only depends on the vacuum pressure $p_\text{vac}$, geometry and the Poisson ratio $\nu$ of the material.
\begin{figure}
\centering
    \includegraphics[width=\columnwidth]{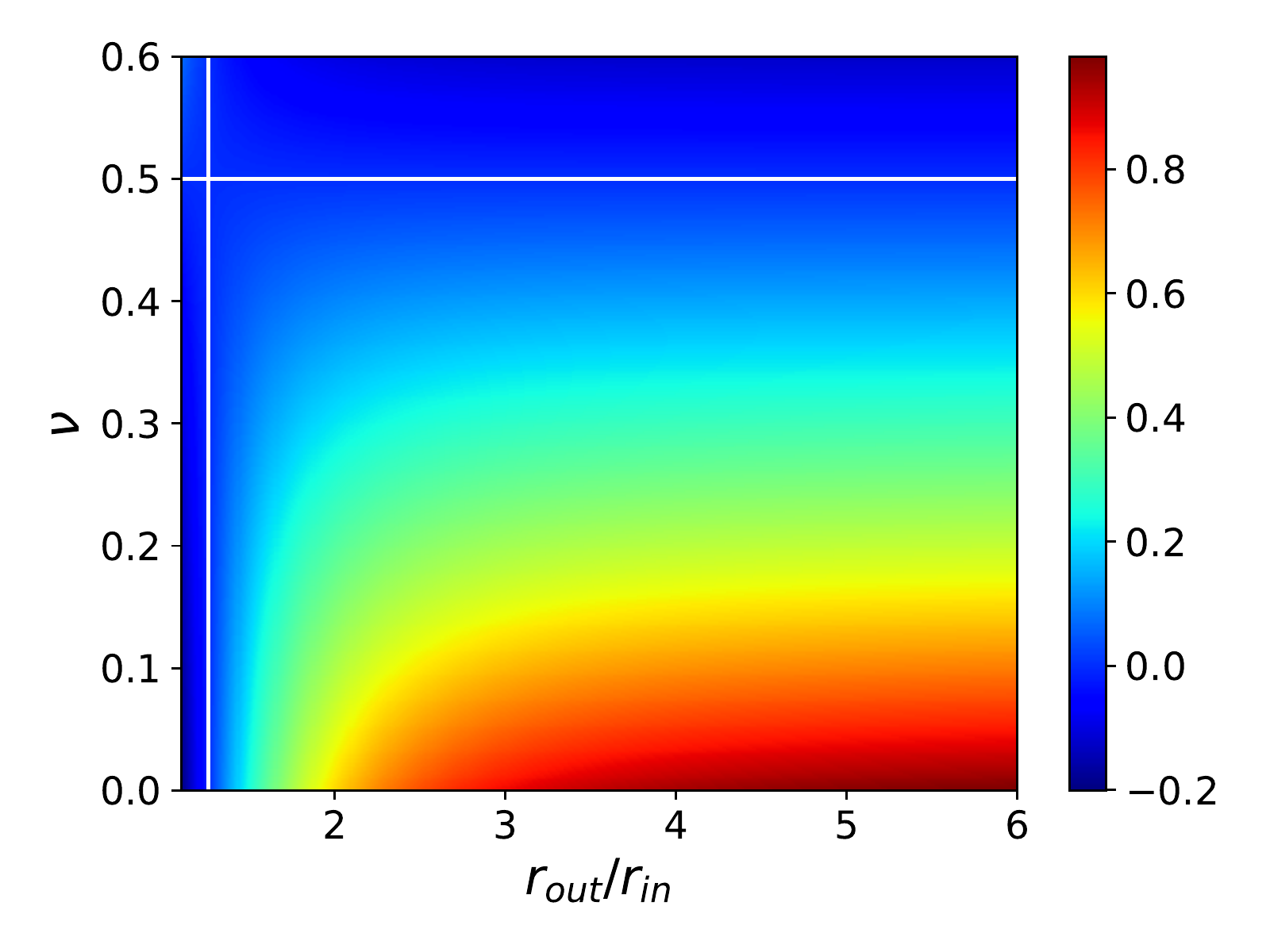}
    \caption{Counterpressure $\frac{p_\text{c}}{p_\text{vac}}$ (color coded) as a function of $\frac{r_\text{out}}{r_\text{in}}$ and the Poisson ratio $\nu$. The white lines indicate where $p_\text{c}$ changes its sign, see text.\label{fig:signOfPc}}
\end{figure}
For $p_\text{c}>0$, the object (inner rigid sphere) gets squeezed by the outer hollow sphere. This mechanism also applies for the hollow hemisphere shown in Fig.~\ref{fig:model} and, hence, for the simple model of the granular gripping system sketched in Fig.~\ref{fig:model}. Fig.~\ref{fig:signOfPc} shows the counterpressure, $p_\text{c}$, as a function of $\frac{r_\text{out}}{r_\text{in}}$ and the Poisson ration $\nu$. The white lines at $\frac{r_\text{out}}{r_\text{in}}=\sqrt[3]{2}$ and $\nu=0.5$ indicate where $p_\text{c}$ changes its sign, and, thus, where the squeezing effect ($p_\text{c}>0$) occurs.
In turn, the squeezing effect causes frictional forces between the membrane of the gripper and the object to be gripped. These frictional forces make it possible to hold and lift the object even in situations where neither interlocking nor vacuum effects between the gripper and the object are possible. In the following, we calculate the gripping force which results from the frictional forces between the object and the membrane.

The force on a small element of surface of the object $\mathrm{d}\vec{A}$ resulting from the compression of the hollow hemisphere is given by $-p_\text{c}\mathrm{d}\vec{A}$. In spherical coordinates, the element of surface reads $\mathrm{d}\vec{A}=r^2\sin\theta\,\mathrm{d}\phi\,\mathrm{d}\theta\hat{e}_r$, where we use the notation defined in Fig.~\ref{fig:hollowSphere}. This force that acts normal to the surface of the object results in a frictional force $-p_\text{c}\mu\mathrm{d}A\hat{e}_\theta$ which is proportional to the normal force and points in negative $\hat{e}_\theta$-direction. Here, $\mu$ denotes the coefficient of friction,
\begin{equation}
\hat{e}_r=
\begin{pmatrix}
\sin\theta\cos\phi\\
\sin\theta\sin\phi\\
\cos\theta
\end{pmatrix}\quad\text{and}\quad
\hat{e}_\theta=
\begin{pmatrix}
\cos\theta\cos\phi\\
\cos\theta\sin\phi\\
-\sin\theta
\end{pmatrix}.
\end{equation}
Adding up both contributions the force on one element of the surface of the object is given by
\begin{equation}
\mathrm{d}\vec{F}=-p_\text{c}\mathrm{d}\vec{A}-p_\text{c}\mu\mathrm{d}A\hat{e}_\theta.
\end{equation}
Thus, the total force on the object reads
\begin{align}
\vec{F} &= -p_\text{c}r_\text{in}^2\cdot\\
&\int_0^\frac{\pi}{2}\mathrm{d}\theta\int_0^{2\pi}\mathrm{d}\phi\sin\theta
\begin{pmatrix}
\sin\theta\cos\phi&+&\mu\cos\theta\cos\phi\\
\sin\theta\sin\phi&+&\mu\cos\theta\sin\phi\\
\cos\theta&-&\mu\sin\theta
\end{pmatrix}
\end{align}
where we integrate the upper hemisphere of the object. The first two components of the integral vanish as $\int_0^{2\pi}\mathrm{d}\phi\sin\phi=\int_0^{2\pi}\mathrm{d}\phi\cos\phi=0$ and for the remaining $z$-component of the force we obtain
\begin{equation}
    \vec{F} = -r_\text{in}^2p_\text{c}2\pi\left(\frac{1}{2}-\mu\frac{\pi}{4}\right)\hat{e}_z\label{eq:fz}.
\end{equation}
To grip the object, $\vec{F}$ needs to point in positiv $z$-direction, i.e. $\mu>\frac{2}{\pi}$. For friction coefficients below the value $\mu_c\equiv\frac{2}{\pi}$, the object will be pressed out of the gripper. For $\mu>\mu_c$ we obtain a gripping force which is proportional to the applied vacuum pressure according to Eq.~\eqref{eq:pc}.

However, the constraint Eq.~\ref{eq:constraint} can also be fulfilled in a trivial way: for $E\rightarrow\infty$, i.e. the case that the hollow sphere itself is made from ideally rigid material, there is no deformation at all, independent on the value of the vacuum pressure $p_\text{vac}$. As there are no materials with infinite Young's modulus this solution seems artificial and not relevant at first glance. However, almost all experiments on granular grippers apply granular particles from rather rigid materials such as glass beads, see e.g. \cite{fitzgerald2020,gomez-paccapeloEffectGranularMaterial2020,lichtUniversalJammingGrippers2016,liSoftRoboticGrippers2019}. But relative to the forces resulting from the vacuum, i.e. the ambient air pressure, dense packings of glass or steel particles may very well be considered as undeformable and rigid. Thus, in this case, the applied vacuum does not lead to the squeezing effect which is present for soft granular particles and the holding forces are reduced. To our knowledge, there is only one publication where particles from soft materials were tested \cite{valenzuela-colomaMentacaUniversalJamming2015}, however, without a detailed discussion of the resulting gripping forces.

\subsection{Particle based simulations of the gripping process}
To explain the gripping force enhancement, we so far assumed that the bulk of the granulate inside the membrane behaves like a linear elastic solid. Of course, a granulate does not obey this simple constitutive law in general. While from a fundamental theoretical point of view the continuum description of granulates is questionable \cite{Goldhirsch:1999}, there are many examples where hydrodynamic descriptions of granular systems yield reliable results, e.g. \cite{goldhirsch:2003,Goldhirsch:2001} and references therein. In these cases, however, other constitutive relations than those of a linear elastic solid are applied, and finding better constitutive laws for granular systems (especially for dense granular flows) is
still a subject of ongoing research e.g. \cite{luding2009towards,jop2006constitutive,barker_gray_2017,schaeffer_barker_tsuji_gremaud_shearer_gray_2019,Kamrin:2015}. To justify our simplifying assumption of a linear elastic solid we, therefore, perform corresponding particle simulations in this section.

\paragraph{Simulation technique}
\begin{figure}
  \centering
  \includegraphics[width=0.95\linewidth]{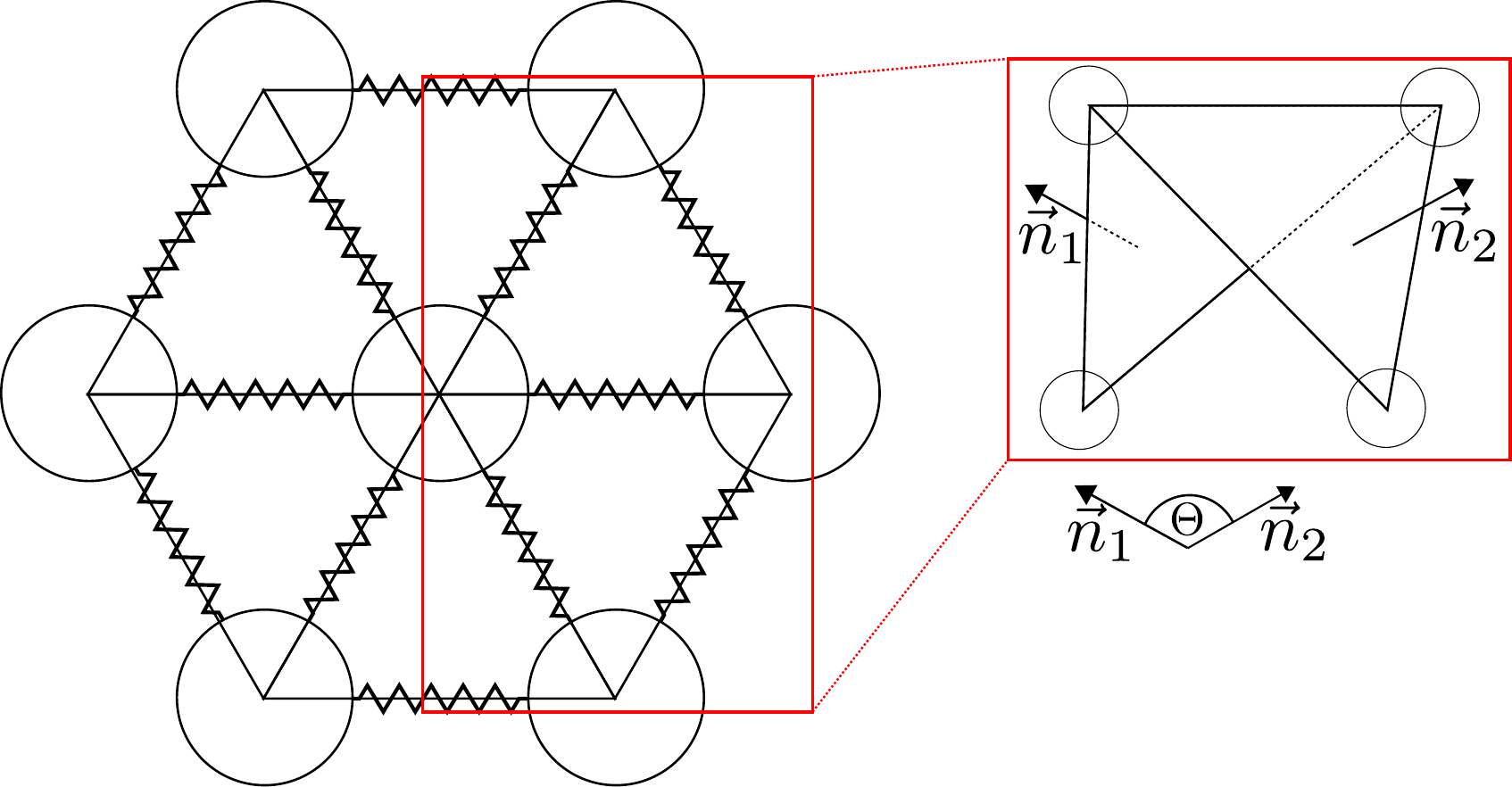}
  \caption{Sketch of the mass spring model of the elastic membrane. The mass particles are ordered on a hexagonal unit cell. Distance springs are indicated. Two neighboring triangle faces enclose the relative angle $\Theta$ defined by the relative direction of their surface normals.}
  \label{img:setup_spring_bending}
\end{figure}

\begin{figure*}
  \centering
  \includegraphics[width=0.24\linewidth]{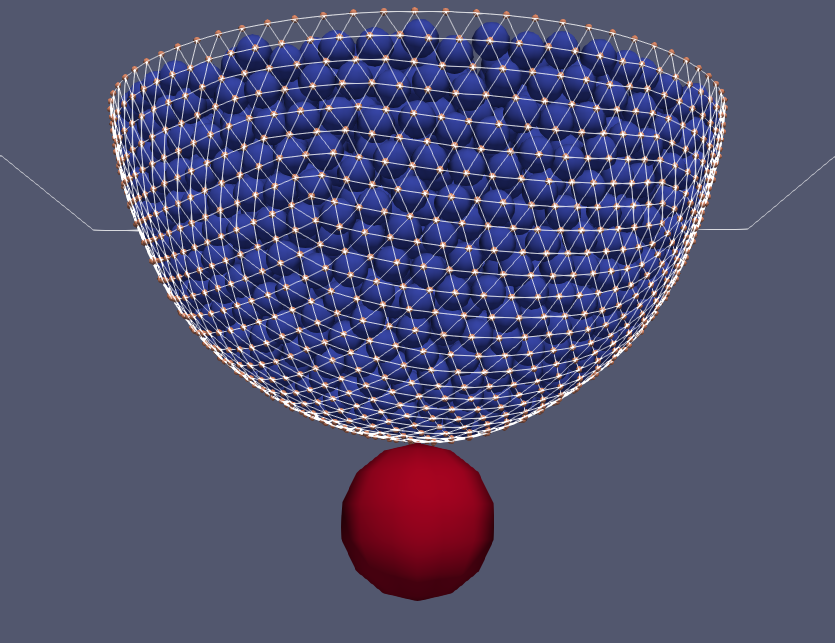}
  \includegraphics[width=0.24\linewidth]{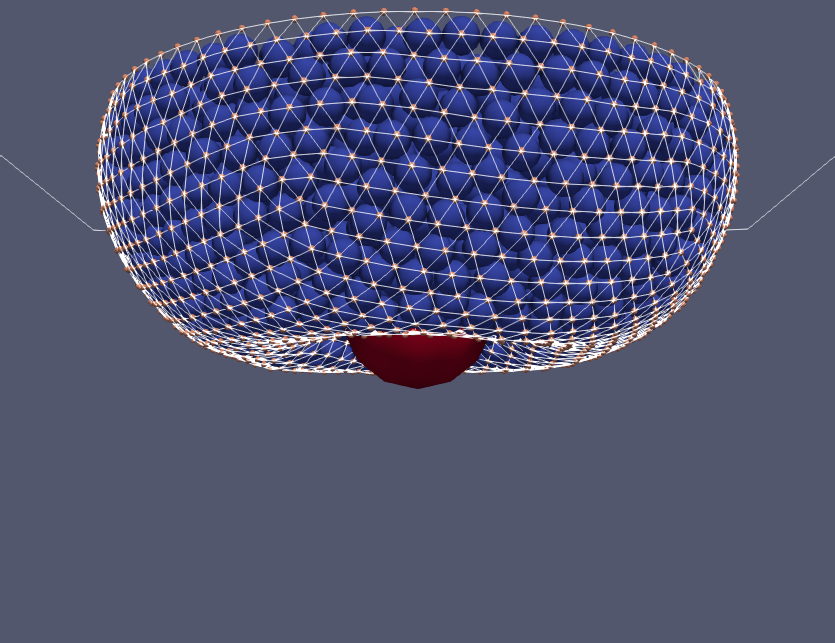}
  \includegraphics[width=0.24\linewidth]{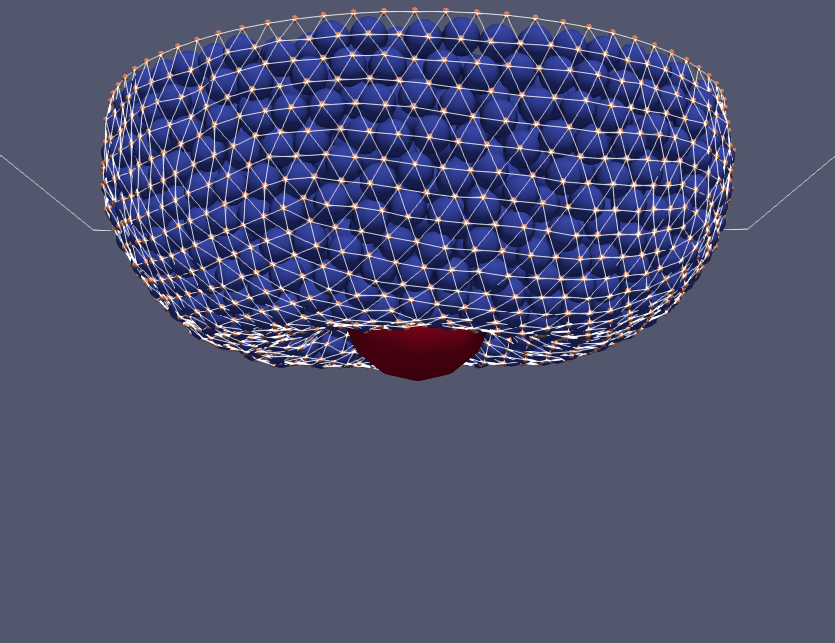}
  \includegraphics[width=0.24\linewidth]{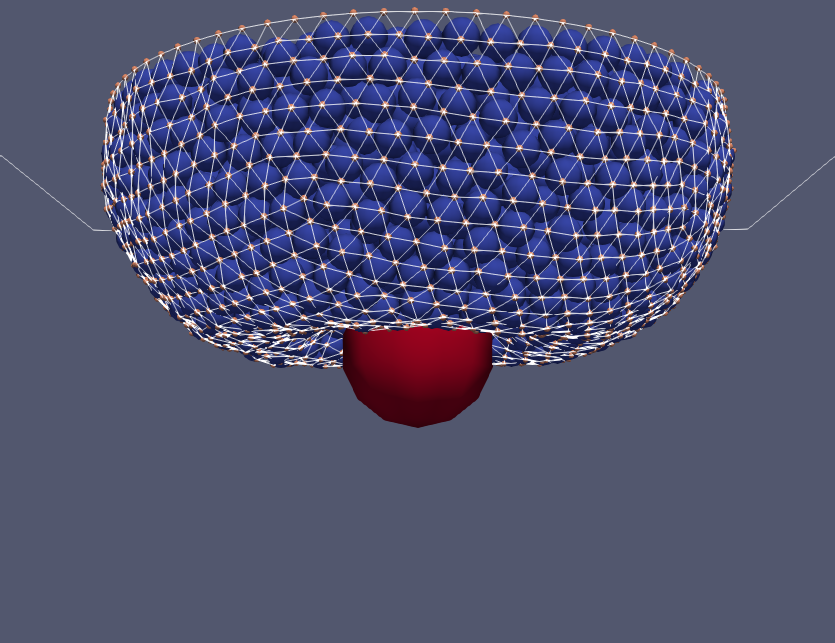}
  \caption{The characteristic phases of the granular gripping process as obtained from the particle simulations. From left to right: initial state, immersion of the object into the granulate, evacuation of the gripper and release of the object.}
  \label{img:sim_procedure}
\end{figure*}

As described in the review article \cite{fitzgerald2020} up to date there is lack of tools for the detailed  simulation of granular gripping processes. In this section we therefore describe a new simulation technique that includes the physics of the granulate on the particle level as well as the mechanics of the membrane and the interaction of the membrane with both, the particles of the granulate and the object.

To simulate the particle dynamics we apply the discrete element method (DEM). The main idea of DEM is to solve Newton's equation of motion to obtain the trajectory, $\vec{r}_i(t)$, for each particle,
\begin{equation}
m_i\frac{\text{d}^2\vec{r}_i}{\text{d}t^2} = \vec{F}_i = \sum\limits_{i=1,j\ne i}^{N} \vec{F}_{ij} + \vec{F}_i^\text{ext}\,,\label{eq:dem_motion}
\end{equation}
where $\vec{F}_i^\text{ext}$ is an external force, such as gravity, acting on the particle and $\vec{F}_{ij}$ is the force acting on particle $i$ due to its contact with particle $j$. A similar equation applies to the rotational degrees of freedom. For a more in-depth introduction of DEM we refer to e.g. \cite{poschelComputationalGranularDynamics2005,matuttisUnderstandingDiscreteElement2014}. Regarding the interactions, a variety of contact models including normal and tangential forces are available. In our work we choose the Hertz-Mindlin no-slip contact model \cite{direnzoComparisonContactforceModels2004}.

In order to simulate the granular gripper's flexible membrane enclosing the granular specimen, we need a method that integrates well with the above described concept of DEM simulations and still yields accurate results. In order to achieve this, we employ a mass spring model which, by design, is easily incorporated into DEM simulations. Additionally, these mass spring models may be used to accurately simulate elastic materials both in 3D and 2D \cite{kotMassSpringModels2017}. Here we chose the computationally cheaper 2D approach, because a thin membrane can be treated as a 2D object to a good approximation.

A mass spring model can be seen as a network consisting of particles carrying mass as vertices and springs as edges, which are used to interconnect the vertices, as shown in Fig.~\ref{img:setup_spring_bending}. This concept can be integrated easily into DEM simulations by adding a force term $\vec{F}_{ij}^\text{spring}$ for the springs to the equation of motion. With this term, which is non-zero only if particles $i$ and $j$ are connected, Equation~\eqref{eq:dem_motion} becomes
\begin{align}
  m_i\frac{\text{d}^2\vec{r}_i}{\text{d}t^2} = \vec{F}_i = \sum\limits_{i=1,j\ne i}^{N} \vec{F}_{ij} + \vec{F}_{ij}^\text{spring} + \vec{F}_i^\text{ext}\,.\label{eq:motion_mass_spring}
\end{align}

With correctly chosen parameters, this concept may be used to represent a membrane, where the interaction between the membrane and the enclosed granular specimen are automatically defined by the contacts between the vertex particles and the granulate. In fact, this approach has been used in recent literature, e.g. \cite{quDiscreteElementModelling2019,debonoDiscreteElementModelling2012}, to simulate triaxial tests. However, this approach may lead to problems. It is, for example, possible that the vertex particles move far apart, due to stretching of the membrane, such that particles from the inside may slip through. More importantly for our application, however, this setup leads to a membrane with a potentially large thickness, depending on the size of the used particles. This would impede a proper simulation of the interaction between the granular gripper and an object that is gripped. Additionally, this approach does not give any smooth surface, which is vital for an accurate simulation of the frictional forces between the membrane and the gripped object. In order to overcome these issues, the vertex particles do not interact with any particle outside the membrane itself. For calculating interactions between the membrane and other particles, we instead use additional wall elements, whose edges are defined by the springs that connect the virtual vertex particles. By tracking the collisions between the vertex particles and the wall elements, this approach additionally allows to detect self intersections of the membrane. To model the bending strength of the membrane, we added additional springs that penalize the bending between two neighboring wall elements. The bending penalty is added as an additional force term to the equation of motion and is quantified by the relative angle between two elements. For a mathematical description of this force, we kindly refer the reader to previous work on cloth simulations done by Bridson et al. \cite{bridsonSimulationClothingFolds2003}. Figure~\ref{img:setup_spring_bending} schematically displays the whole setup together with the definition of the angle. The dynamics of the membrane are mapped to the following steps for each time step iteration of the DEM simulation:
\begin{enumerate}
    \item The position of the wall elements is updated according to the positions of the vertex particles. Similarly, the velocity is passed on to the wall elements as it is needed for the  calculation of the (frictional) contact forces.
    \item Contacts between wall elements and other objects in the simulation are detected and computed using the Hertz-Mindlin non slip model.\label{enum:wall_forces}
    \item The forces acting on the wall elements are distributed to the surrounding vertex particles according to the bary\-centric coordinates of the contact point.
    \item Additional forces due to the angle between adjacent wall elements and the displacement of the springs are calculated and added to the node particles.
    \item The system is integrated and the positions and velocities of the vertex particles are updated.
\end{enumerate}
The vacuum, which is vital for the gripping process, causes a pressure difference $\Delta p$ between the inside and the outside of the membrane which, in turn, leads to forces acting on the membrane. In our model, these forces are determined on a per wall element basis: The magnitude is calculated by multiplying the elements surface area with the pressure difference $\Delta p$. The direction is determined by the element's inwards pointing normal vector. The resulting forces are then applied in step \ref{enum:wall_forces}. This procedure reproduces the correct force magnitude when compared to reality, merely the force direction is expected to deviate locally, due to the flat wall element's finite size.

\begin{figure}
\centering
    \includegraphics[width=0.95\columnwidth]{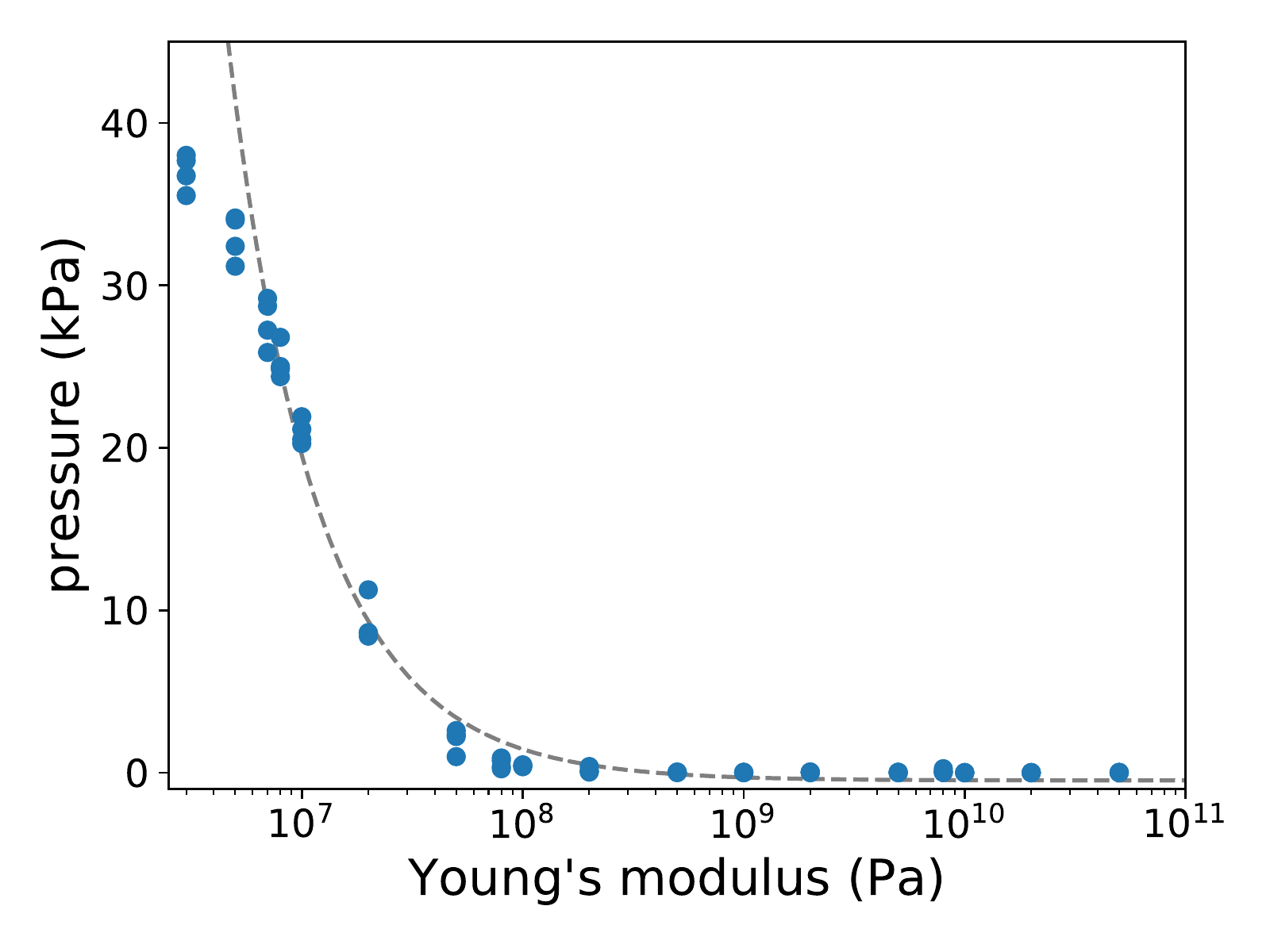}
    \caption{Pressure onto the surface of rigid sphere inside a granular gripper according to the model setup shown in Fig.~\ref{fig:model}. The blue circles show the data obtained from particle simulations, the dashed grey line shows a hyperbolic fit to the data.\label{fig:pSim}}
\end{figure}

\begin{figure}
\centering
    \includegraphics[width=0.95\columnwidth]{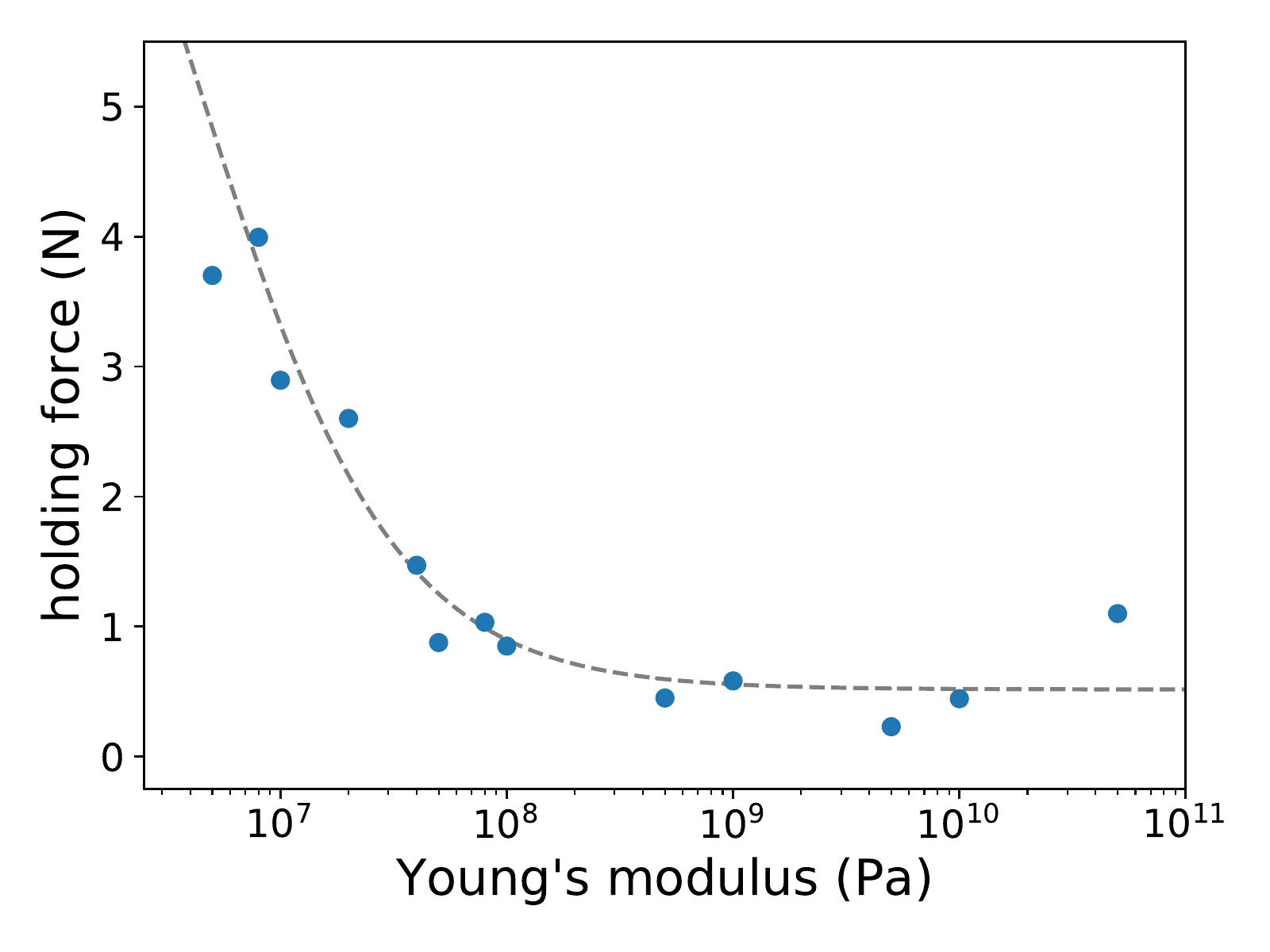}
    \caption{Maximal holding force for the granular gripping process shown in Fig.~\ref{img:sim_procedure} as a function of the stiffness of the material of the granular particles. The blue circles show the data obtained from particle simulations, the dashed grey line shows a hyperbolic fit to the data.\label{fig:fSim}}
\end{figure}

Using this setup we were able to simulate the gripping process. Figure~\ref{img:sim_procedure} shows simulation snapshots for the different phases of the gripping process. First, we reconsider the model scenario of the pressurized hollow sphere from the previous section (see Fig.~\ref{fig:hollowSphere}). However, we now replace the linear elastic material by two spherical membranes of radius $r_\text{in}\approx0.02\,\text{m}$ and $r_\text{out}\approx0.034\,\text{m}$. \red{Similar to the inner membrane also the inner object has a radius of $0.02\,\text{m}$.} The membranes are represented by 1280 triangles each, and the stiffness constant $k$ for the springs that interconnect the vertices of the triangles is chosen to be $k=324.76\frac{\text{N}}{\text{m}}$, such that a membrane of thickness $t=0.3\,\text{mm}$, an elastic modulus \red{$E=\frac{2\sqrt{3}}{3t}k=1.25\cdot10^{6}\,\text{Pa}$} and a Poisson ratio of $\nu=0.33$ is approximated \cite{kotMassSpringModels2017}. The  space between the two membranes is filled with particles up to random close packing. The parameters of the simulated particles are given in Tab.~\ref{tab:matProp}.
\begin{table}
\centering
\begin{tabular}{lr}
 property & value\\\hline
 radius [m] & $0.002$\\
 elastic modulus [Pa] & between $10^{6}$ and $10^{10}$\\
 Poisson's ratio & 0.245\\
 coefficient of restitution & 0.5\\
 friction coefficient & 0.16\\
\end{tabular}
\caption{Parameters used for the simulation\label{tab:matProp}}
\end{table}
Now a pressure gradient $\delta p = 90\,\text{kPa}$ is applied in the simulation. Fig.~\ref{fig:pSim} shows the resulting pressure on a rigid spherical object inside the hollow sphere. If the stiffness of the material of the particles is larger than approximately $10^8\,\text{Pa}$, the bulk of the granulate may be considered as a rigid material and the pressure vanishes as described for the case of a linear elastic material in the previous section. If the stiffness $E$ of the particle material decreases below $10^8\,\text{Pa}$, the pressure increases as $\frac{1}{E}$. This again coincides with the theoretic prediction for a linear elastic solid instead of the granulate (see Eq.~\eqref{eq:dissplacement}). In our simulations, we assume spherical particles and overlaps of the spheres are interpreted as mutual deformations of the spheres, which, in turn, cause repulsive forces. This interpretation of the overlap is only valid for overlaps, which are small compared to the size of the particles. If the stiffness of the particles material is bellow $\approx 5.0\cdot10^6\,\text{Pa}$ this is no longer valid and our simulations results are not reliable anymore. We therefore limit Fig.~\ref{fig:pSim} (and the later Fig.~\ref{fig:fSim}) to elastic moduli above $E\approx 5.0\cdot10^6\,\text{Pa}$.

Next, we determine the maximal holding force for the granular gripping process shown in Fig.~\ref{img:sim_procedure}. For this purpose we fill a spherical bag of radius $r=0.04\,\text{m}$ with granulate, evacuated it and thereby grip a spherical object. The parameters of the membrane and the particles are the same as for the hollow sphere simulations. Additionally, the membrane's friction coefficient is set to $1.16$ and the particles' coefficient of restitution is changed to $0.926$ to mimic the material properties of latex. The parameters of the gripped object are given in Tab.~\ref{tab:simProb}.
\begin{table}
\centering
\begin{tabular}{lr}
 property & value\\\hline
 radius [m] & $0.01$\\
 elastic modulus [Pa] & $2.8\cdot 10^{9}$\\
 Poisson's ratio & 0.38\\
 coefficient of restitution & 0.67\\
 friction coefficient & 1.2\\
\end{tabular}
\caption{Simulation parameters of the gripped sphere}
\label{tab:simProb}
\end{table}
Corresponding to the linear elastic theory of the previous section and in agreement with the course of the pressure curve shown in Fig.~\ref{fig:pSim}, we again identify the limit where the bulk of the granulate behaves ideally rigid for Young's modulus $E>10^8\, \text{Pa}$, see Fig.~\ref{fig:fSim}. In this regime, the squeezing effect and, correspondingly, the holding force, is small. Note that unlike the case where a linear elastic solid is considered instead of the granulate, the holding force is small but does not vanish. Similarly to the pressure, for Young's modulus $10^7\, \text{Pa}<E<10^8\, \text{Pa}$, the holding force increases like $1/E$.


\section{Summary and Outlook}
Most research on application examples for granular grippers use particles of rather rigid material. This approach is obvious, as for stiff particles the gripper hardens well, from which a strong gripping performance is expected. However, this only works well in situations where geometric interlocking between the object and the gripper or suction effects between the membrane and the object are possible. If this is not possible due to the shape of the object or its surface properties, friction is the only way to build up significant holding forces. Significant friction however, is only possible, if the membrane is distinctly pressed onto the object. In this work we have shown that this is not possible if rigid granular particles are used. By experiments, theory and particle simulations we have revealed that using soft instead of almost rigid particles unavoidably leads to a squeezing effect and, hence, to significant friction and holding forces. In the course of that, we developed a new and promising method for the simulation of granular gripping processes.

Apparently, the holding force enhancement due to soft particles is limited: On the one hand, the soft particles increase the holding force, but on the other, they make the gripper deformable in its entirety and a trade off between both effects is necessary.

\begin{acknowledgements}
H.G., A.S. and T.P. gratefully acknowledge funding by the Deutsche Forschungsgemeinschaft (DFG, German Research Foundation) – project number 411517575. P.M. gratefully acknowledges funding by the DFG under project number 398618334. All the authors thank W. Pucheanu for his contribution in the design and construction of the experimental setup. The work was also supported by the Interdisciplinary Center for Nanostructured Films (IZNF), the Central Institute for Scientific Computing (ZISC), and the Interdisciplinary Center for Functional Particle Systems (FPS) at Friedrich-Alexander University Erlangen-N\"urnberg.
\end{acknowledgements}

%
%


\end{document}